\RequirePackage{fix-cm}
\documentclass[smallextended]{svjour3}       
\smartqed  
\usepackage{graphicx}

\usepackage{fancyhdr}
\usepackage{array}
\usepackage{graphicx}
\usepackage{cite}
\usepackage{amsmath}
\usepackage{color}
\begin{document}

\title{Collectivity and isomers in the Pb isotopes}

  \author{  Praveen C. Srivastava$^{1,*}$  \and Sakshi Shukla$^{1}$ }

\institute{ *Email: praveen.srivastava@ph.iitr.ac.in
\at
Department of Physics, Indian Institute of Technology, Roorkee - 247667, INDIA\\
 }

\date{Received: date / Accepted: date}

\maketitle

\begin{abstract}

In the present work, we aim to study collectivity in the Pb isotopes in the framework of nuclear
shell model. We have performed shell-model calculations using  KHH7B effective
interaction. The model space of KHH7B interaction consists
of 14 orbitals. We have reported 
results for even-even $^{196-206}$Pb isotopes for spectra and electromagnetic properties. The shell model results for isomeric states are also reported. Our results will be useful to compare
upcoming experimental data.\\
\end{abstract}

{\bf Keywords:} Collectivity, Isomers, Shell-Model

\section{Introduction}\label{sec1}

The investigation of Pb isotopes around N=126 has been a crucial area for the experimental and theoretical point of view \cite{Brown_PRL,Bhoy2,pcs}. Furthermore, there exists an extensive research focused on the structure and electromagnetic properties of the spherical Pb isotopes\cite{blomqvist}.
Due to the shell closure at Z = 82, these nuclei exhibit spherical configurations in their ground states. However, the interplay between single-particle motion, collective behavior, and pairing interactions has unveiled a diverse array of coexisting structures at relatively low energy levels\cite{wood}. These coexisting structures can present different shapes, giving rise to
a phenomenon known as the shape coexistence\cite{heyde}. Comprehensive investigations into $\alpha$-decay and $\beta^+$/electron capture-decay have provided substantial evidence for the occurrence of low-lying excited $0^+$ states in all even-even Pb isotopes with mass numbers varying from A = 184 to A = 200\cite{andreyev,julin}. {\color{black}  Study of $^{198}$Pb isotope has been done by performing Coulomb excitation experiment using the miniball $\gamma$ ray spectrometer and radioactive ion beam from the REX-ISOLDE post-accelerator at CERN\cite{pakarinen}.} The competition between allowed and forbidden beta decays
in this region put additional challenges for theoretical model to predict more precisely wave functions for further study of weak interaction processes \cite{carroll2020,brunet2021}.
Thus, motivated by these findings we aim to study energy spectra, electromagnetic transitions and nuclear isomers for Pb isotopes.

\section{Results and Discussions}\label{sec2}

The nuclear shell-model Hamiltonian can be expressed in terms of
single-particle energies and two nucleon interactions,
\begin{equation}
H = \sum _{\alpha}\epsilon _{\alpha}c_{\alpha}^{\dagger}c_{\alpha}+
\frac{1}{4}\sum _{\alpha \beta \gamma \delta JT}\langle j_{\alpha }j_{ \beta
}|V|j_{\gamma }j_{\delta} \rangle _{JT}c_{\alpha}^{\dagger}c_{
\beta}^{\dagger}c_{\delta }c_{\gamma },\label{eq1}
\end{equation}
where $\alpha =\{n,l,j,t\}$ stand for the single-particle orbitals and
$\epsilon _{\alpha}$ are corresponding single-particle energies. 
The $c_{\alpha}^{\dagger}$ and $c_{\alpha}$ are the fermion annihilation and creation operators.  The antisymmetrized two-body matrix elements coupled to spin
$J$ and isospin $T$ is given by $\langle j_{\alpha }j_{\beta }|V|j_{\gamma }j_{\delta }\rangle
_{JT}$ denote. To diagonalize the matrices the shell-model code KSHELL \cite{KSHELL} has been employed. {\color{black} In the present work we have performed shell-model study of $^{196-206}$Pb isotopes using KHH7B \cite{KHH7B,Kuo,mcgrory} effective interactions.} The KHH7B interaction consists of 14 orbitals having the proton orbitals between $Z = 58-114$: $1d_{5/2}$, $1d_{3/2}$, $2s_{1/2}$, $0h_{11/2}$, $0h_{9/2}$, $1f_{7/2}$, $0i_{13/2}$ and neutron orbitals between $N = 100-164$: $1f_{5/2}$, $2p_{3/2}$, $2p_{1/2}$, $0i_{13/2}$, $1g_{9/2}$, $0i_{11/2}$, $0j_{15/2}$. {\color{black} In the present calculation, we have allowed neutrons to occupy between $N$=100-126 shell i.e. in $1f_{5/2}$, $2p_{3/2}$, $2p_{1/2}$ and $0i_{13/2}$ orbitals, while protons orbitals ($1d_{5/2}$, $1d_{3/2}$, $2s_{1/2}$ and $0h_{11/2}$)  are completly filled up to $Z=82$.}

\begin{figure*}
\begin{center}
\includegraphics[width=9.7cm,height=7.5cm,clip]{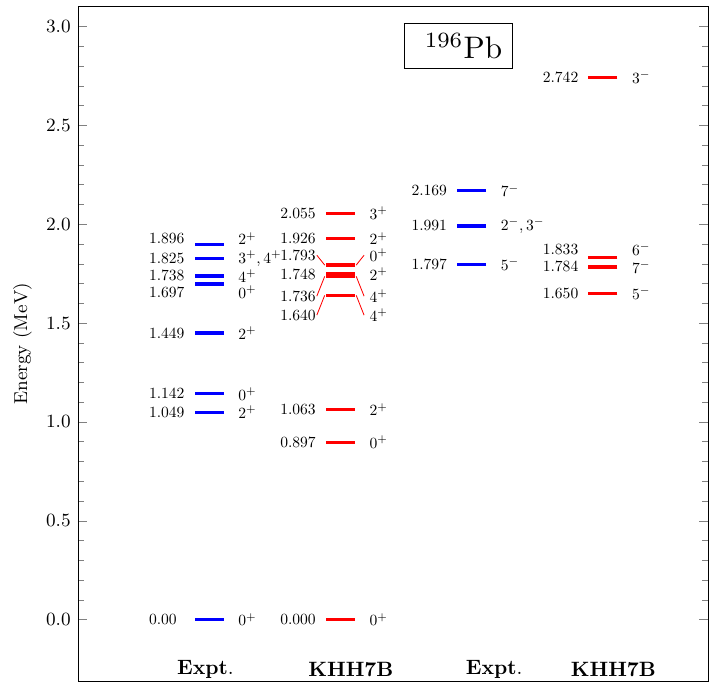}
\includegraphics[width=9.7cm,height=7.5cm,clip]{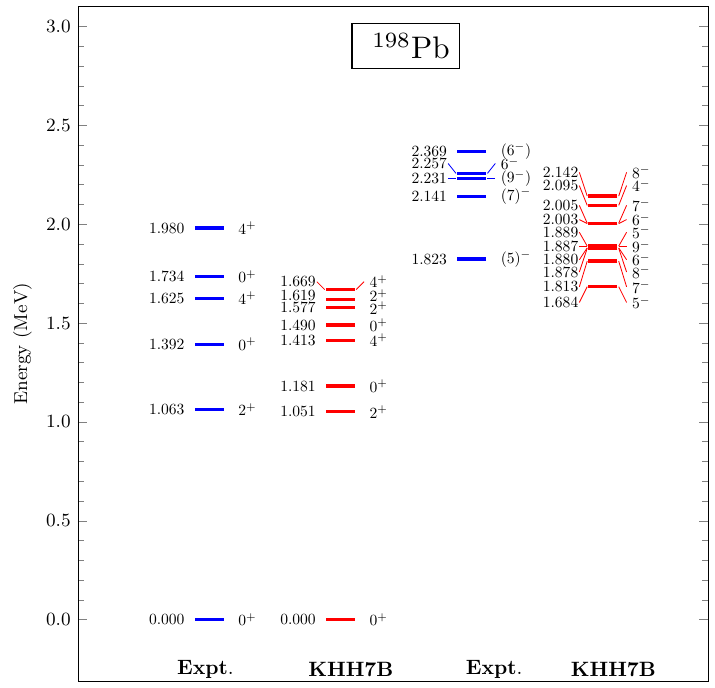}
\includegraphics[width=9.7cm,height=7.5cm,clip]{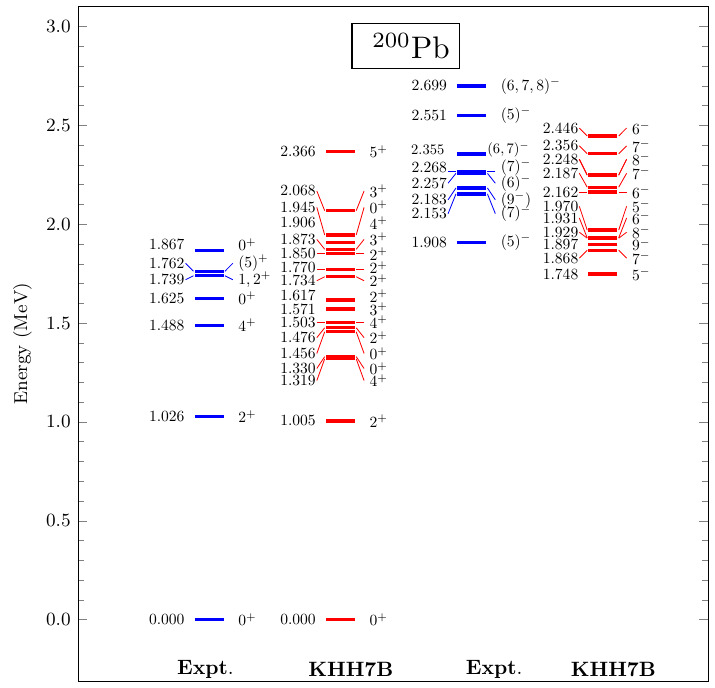}
\caption{\label{spectra1}Comparison between experimental\cite{nndc} and calculated energy levels for $^{196-200}$Pb isotopes.}
\end{center}
\end{figure*}

\begin{figure*}
\begin{center}

\includegraphics[width=9.7cm,height=7.5cm,clip]{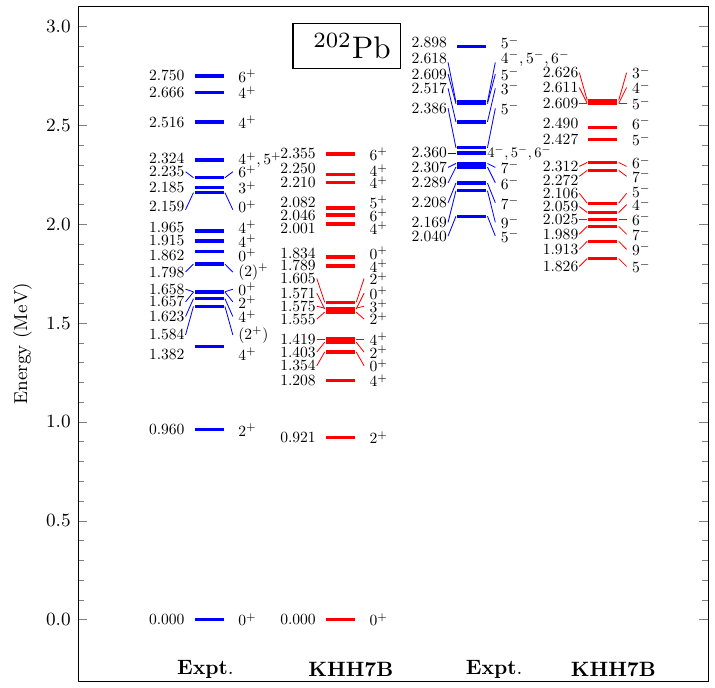}
\includegraphics[width=9.7cm,height=7.5cm,clip]{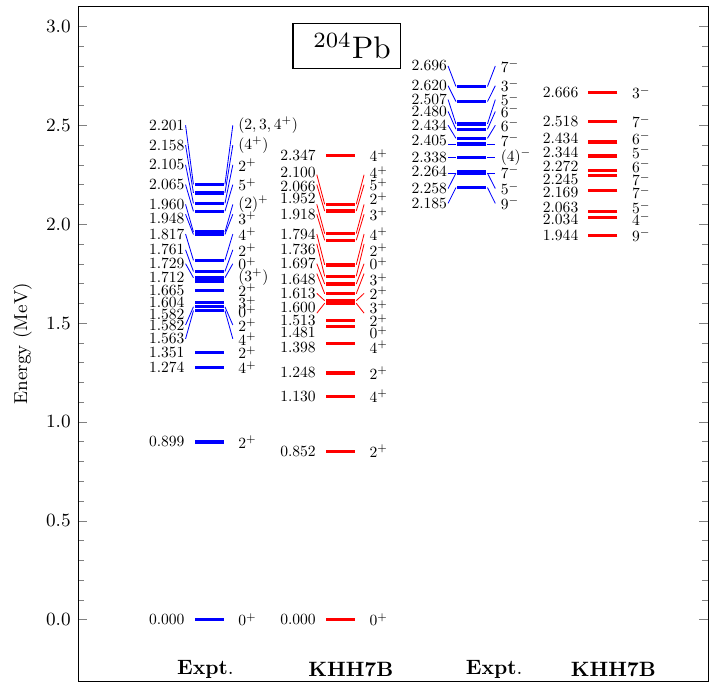}
\includegraphics[width=9.7cm,height=7.5cm,clip]{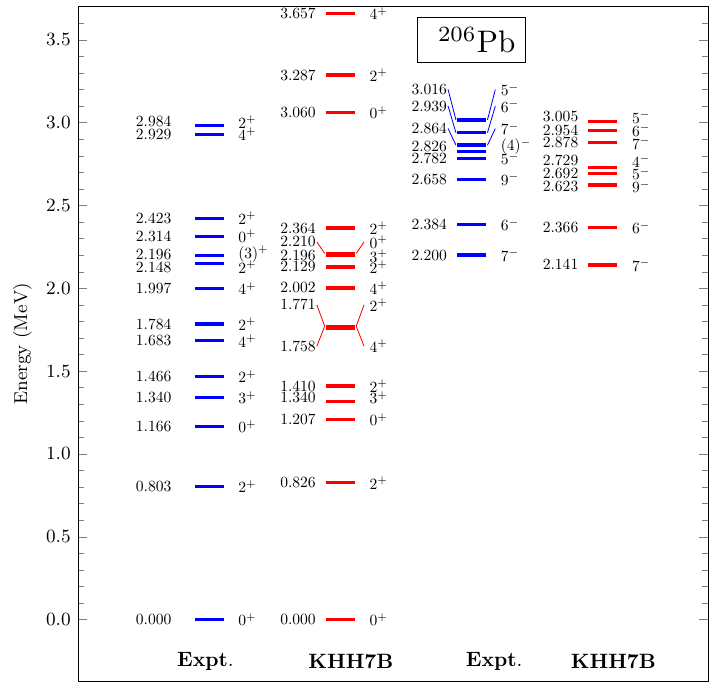}
\caption{\label{spectra2}Comparison between experimental\cite{nndc} and calculated energy levels for $^{202-206}$Pb isotopes.}
\end{center}
\end{figure*}

Figs. \ref{spectra1}-\ref{spectra2} shows a comparison between shell-model results and experimentally observed levels for $^{196-206}$Pb isotopes. Here we have calculated only low-lying states. Our calculated result shows reasonable agreement with the experimental data. Yrast $0^+$, and $2^+$ states are reproduced very well for all the even-even Pb isotopes with mass numbers ranging from 196-206. All the negative parity states are compressed.
There are minimum of three $0^+$ states in all $^{196-206}$Pb, all of these states are lying below 2.0 MeV, except $^{206}$Pb. The 2$^+$ states in $^{196-198}$Pb are formed due to the two-quasineutron excitations to the i$_{13/2}$ orbital. We see a large energy difference between $0^+$-$2^+$ states and then $2^+$-$4^+$ states while all other states are lying close for $^{196-206}$Pb isotopes.
For $^{196-202}$Pb isotopes lowest lying negative parity state in these isotopes is $5^-$, which is also reproduced theoretically. In $^{196-200}$Pb isotope the $5^-$ state is coming from the coupling of $p_{3/2}$ and $i_{13/2}$ orbital, and in $^{202}$Pb isotope its coming due to $f_{5/2}$ and $i_{13/2}$ orbitals.
For $^{196}$Pb isotope experimentally at 1.825 MeV excitation energy, $3^+$, and $4^+$ states are degenerate, while theoretically, we see a large differences in their excitation energy. Among the three $0^+$ states, $0_1^+$, and 0$^+_3$ states are reproduced well.
The calculated collective $3^-$ state also show much discrepancy with its experimental data, probably with the shell-model it is possible to reproduce if we take extended model space \cite{Brown_PRL}.

{\color{black} In $^{198}$Pb isotope we get large energy difference between $0^+_1$, and $0^+_2$ state theoretically, also for the $0^+_2$, and $0^+_3$ states the energy gap decreases.} All the negative parity states are compressed very much. The $6^-_1$, and $7^-_1$ states are lying close theoretically, as well as observed in experimental data, formed from the same configuration $\nu (f_{5/2}i_{13/2})^{-1}$.

The $^{200}$Pb spectra above 1.4 MeV are compressed except for the $(5)^+$ state at 1.762 MeV excitation energy. In this isotope all the negative parity states are tentative. {\color{black} Experimentally at 2.355 MeV excitation energy degenerate state is $(6,7)^{-1}$ obtained, whereas theoretically, these states are lying at different energy.} 
{\color{black} The experimental $(5)^-$, $(7)^-$, and $(9^-)$ states are at 1.908, 2.153, and 2.183 MeV excitation energy, while calculated values of these $5^-$, $7^-$, and $9^-$ states
are 1.748, 1.868, and 1.897 MeV, respectively.}
Experimentally at 2.268 MeV excitation energy $(7)^{-}$ state with unconfirmed spin is obtained, but theoretically, this state is lying 81 keV below, so we can predict the spin to be 7.

For $^{202}$Pb isotope up to 2.208 MeV excitation energy all the yrast states are reproduced in the same order as experimental level but compressed. 
The yrast $0^+$, $2^+$, and $4^+$ states are formed due to $\nu (f_{5/2})^{-2}$ configuration. We observe large separation between yrast $4^+$, and $6^+$ states both theoretically as well as experimentally. Experimentally at 2.360 MeV excitation energy $4^-$, $5^-$, and $6^-$ states are degenerate, whereas theoretically, these states are non-degenerate, formed due to $f_{5/2}$, and $i_{13/2}$ orbitals.  
The low-lying spectra of $^{200-204}$Pb show much similarity.

{\color{black} In the $^{204}$Pb isotope, the unconfirmed experimentally ($3^+)$ state at 1.712 MeV corresponds to the calculated $3_2^+$ state obtained at 1.648 MeV.}
Similarly, our result supports the unconfirmed experimentally ($2^+$), and ($4^+$) levels at 1.960, and 2.158 MeV excitation energy to be $2^+_6$, and $4^+_4$, respectively by comparing both of these states with their theoretical result. 
{\color{black} The shell-model calculated values for $2^+_6$ and  $4^+_4$ states are 1.952, 2.108 MeV, respectively.}
For $^{206}$Pb isotope positive parity states up to 2.5 MeV excitation energy, and all the negative parity states are reproduced very well. {\color{black} The calculated spectrum is compressed for
$^{196-204}$Pb, while for $^{206}$Pb energy levels are in very good agreement this is because 
once we move towards shell-closure shell-model predictive power is better}.

\begin{figure}
\begin{center}
   \includegraphics[width=5.7cm,height=5cm,clip]{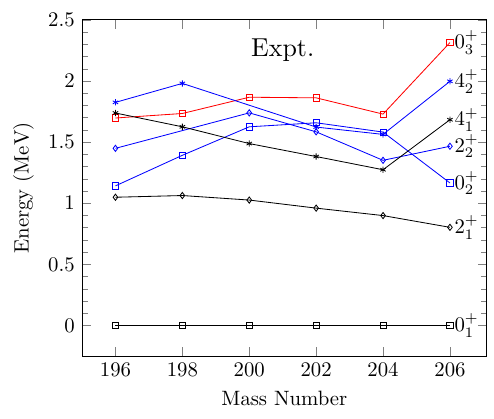}
   \includegraphics[width=5.7cm,height=5cm,clip]{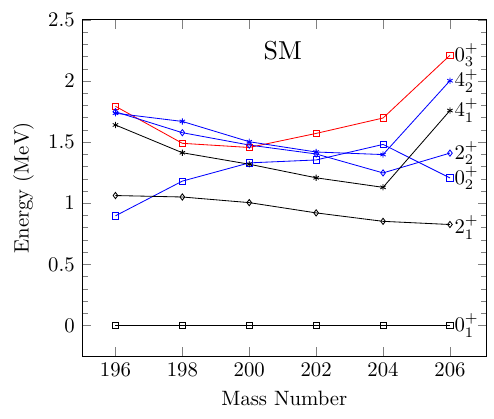}
   \caption{Level-energy systematics of the neutron-deficient Pb isotopes with $A=196-206$.}
   \label{2+}
  \end{center} 
\end{figure}

In the case of $^{206}$Pb, the magnetic moment of the 6$^-_1$ state was found to be positive, and that of the 7$^-_1$
state to be negative in experimental observations. However, in the current calculations, these values are determined to be negative. Both the 6$^-_1$ and 7$^-_1$ states share the $\nu (p_{1/2}^{-1}i_{13/2}^{-1})$ configuration, and theoretically, both states should exhibit the same sign.
In Fig. \ref{2+}, the comparison between calculated 
$0_1^+$-$2_1^+$-$0_2^+$-$2_2^+$-$4_1^+$-$4_2^+$-$0_3^+$
 is shown. The shell-model results show a reasonable trend in comparison to the experimental data. 


The isomeric state $5^-$, and $9^-$ in $^{196}$Pb isotope consists configuration $\nu (p_{3/2}^{-1}i_{13/2}^{-1})$, and $\nu (f_{5/2}^{-1}i_{13/2}^{-1})$, respectively, thus their seniority quantum number ($v$) is 2. The isomeric $5^-$ state in $^{198}$Pb isotope is formed due to coupling of $p_{3/2}$ and $i_{13/2}$ orbital thus seniority ($v$) is 2. The isomeric states $7^-$, and $9^-$ consists configuration $\nu (f_{5/2}^{-1}p_{3/2}^{2}i_{13/2}^{-1})$, are formed due to one unpaired neutron hole in $f_{5/2}$, and $i_{13/2}$ orbital each thus seniority is 2. The $7^-$, and $9^-$ isomers in $^{200}$Pb consists of configuration $\nu (f_{5/2}^{-3}p_{3/2}^{2}i_{13/2}^{-1})$, and obtained due to one unpaired neutron hole in $f_{5/2}$, and $i_{13/2}$ orbital each. Therefore seniority of both these states is 2. Like other isotopes discussed above $^{202}$Pb isotope also have $7^-$, and $9^-$ as isomeric state having configuration $\nu (f_{5/2}^{-3}i_{13/2}^{-1})$ and coming from one unpaired neutron hole in $f_{5/2}$, and $i_{13/2}$ orbital. Thus both of these states have seniority quantum number ($v$) = 2.
The $4^+$ isomeric state in $^{204}$Pb is formed due to one neutron hole pair in $f_{5/2}$ orbital, therefore seniority of this state is 2. Alike $7^-$, and $9^-$ isomers discussed above $7^-$ [$\nu (f_{5/2}^{-2}p_{1/2}^{-1}i_{13/2}^{-1})$, and $9^-$ [$\nu (f_{5/2}^{-1}i_{13/2}^{-1})$] isomer observed in $^{204}$Pb isotope does not have similar configuration. Among these isomers $7^-$ is formed due to one unpaired neutron in $p_{1/2}$, and one unpaired neutron hole in $i_{13/2}$ orbital each, whereas $9^-$ is obtained from one neutron hole in $f_{5/2}$, and $i_{13/2}$ orbital each. Therefore both of these states have seniority ($v$) equal to 2. The $7^-$ isomeric state in $^{206}$Pb isotope formed due to coupling of one neutron hole in $p_{1/2}$, and $i_{13/2}$ orbital each, thus seniority ($v$) is 2.
Hence all low lying isomeric states obtained for $^{196-206}$Pb isotope have maximum seniority quantum number 2. We have also calculated half-lives of those isomeric states that undergoes E2 transition, using $B(E2)$ value. We have reproduced the half-lives of $7^-$ isomeric state in $^{198}$Pb very well i.e. 1.2 $\mu s$ (corresponding experimental value is 4.19(10) $\mu s$). 
{\color{black} The calculated half-life half-life for $9^-$ isomeric state in $^{198}$Pb isotope is 4199 $n s$, while corresponding experimental value is 137(10) $ns$.}
We have nicely reproduced the half-life for the $7^-$ isomeric state in $^{204}$Pb isotope, which is 0.40 $\mu s$ (corresponding experimental value is $0.45^{+10}_{-3} \mu s$). 
{\color{black} The calculated half-life of the $4^+$ isomeric state in $^{204}$Pb isotope is 36.2 $n s$, while corresponding experimental value is 265(6) $ns$.}
We see that the $7^-$ is an isomeric state in $^{198-206}$Pb. It is coming from one unpaired neutron in $f_{5/2}$ and $i_{13/2}$ each, except for the $^{204}$Pb. The $9^-_1$ is an isomeric state in all $^{196-204}$Pb and is formed due to one unpaired neutron in $f_{5/2}$, and $i_{13/2}$ orbital each, with maximum probability of 68.18 $\%$ for $^{204}$Pb and minimum probability of 29.78 $\%$ for $^{198}$Pb.

\begin{figure}
\begin{center}
   \includegraphics[height=5.3cm,clip]{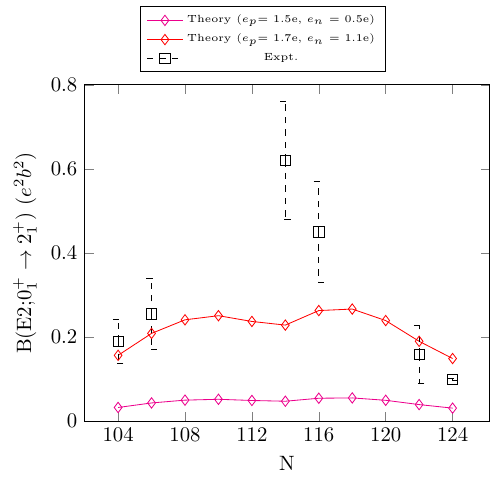}
    \includegraphics[height=5.1cm,clip]{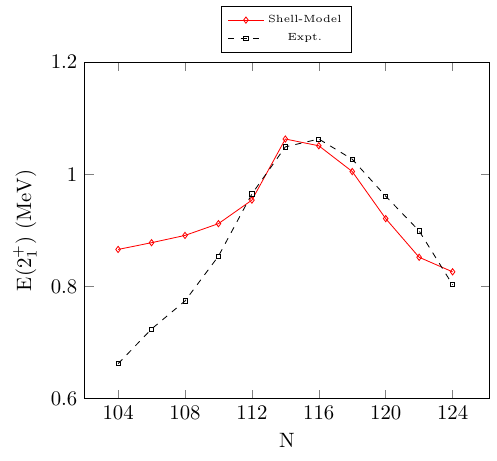}
   \caption{ Comparison between calculated and experimental\cite{nndc,nudat}  B(E2;0$^+_1$ $\rightarrow$ 2$^+_1$) for even-even $^{186-206}$Pb in the left panel. Calculated and experimental $E(2_1^+$) energy is also shown in the right panel.}
   \label{be2_trend}
   \end{center}
\end{figure}


\begin{table*}
\caption{\label{be2} The calculated $B(E2)$ values in units of W.u. for Pb
isotopes using KHH7B interaction (SM) compared to the experimental data (Expt.)
\cite{nndc,pakarinen} corresponding to $e_p$ = 1.5$e$ and $e_n$ = 0.5$e$.  }

\begin{tabular}{rrc|cccc}
\hline      
& ${B(E2; J_i \rightarrow   J_f}$)  & ~~~~~ &  ~~~~~ ${B(E2; J_i \rightarrow   J_f}$) &      &\\
\hline
\hline
& &    &   &     &    \\
$^{196}$Pb & Expt. & SM & $^{198}$Pb  & Expt. & SM\\
\hline
 {2}$^+_1$ $\rightarrow$ {0}$^+_1$&18.2$_{-4.1}^{+4.8}$ &1.4& {2}$^+_1$ $\rightarrow$ {0}$^+_1$&13.1$_{-3.5}^{+4.9}$ &1.6\\
{4}$^+_1$ $\rightarrow$ {2}$^+_1$&NA & 3.3$\times 10^{-7}$  &{4}$^+_1$ $\rightarrow$ {2}$^+_1$&NA &0.4    \\
{9}$^-_1$ $\rightarrow$ {7}$^-_1$& $1.17(12)$ &0.1    & {7}$^-_1$ $\rightarrow$ {5}$^-_1$&0.000553(14) &0.1    \\
& & &{9}$^-_1$ $\rightarrow$ {7}$^-_1$&0.93(7)&3.5$\times 10^{-2}$\\
\hline
& &    &   &     &    \\
$^{200}$Pb & Expt. & SM & $^{202}$Pb  & Expt. & SM\\
\hline
 {2}$^+_1$ $\rightarrow$ {0}$^+_1$&NA &1.6&{2}$^+_1$ $\rightarrow$ {0}$^+_1$&$>$0.098 &1.4\\
 4$^+_1$ $\rightarrow$ 2$^+_1$  & 1.13(7)    & 0.2&{4}$^+_1$ $\rightarrow$ {2}$^+_1$&0.291(21) &5.0$\times 10^{-2}$ \\
 7$^-_1$ $\rightarrow$ 5$^-_1$  & 0.167(3)    & 4.7$\times 10^{-3}$ & {0}$^+_2$ $\rightarrow$ {2}$^+_1$&$>$1.4 &0.1   \\
{9}$^-_1$ $\rightarrow$ {7}$^-_1$&0.38(10) &1.0$\times 10^{-2}$& {0}$^+_3$ $\rightarrow$ {2}$^+_1$&$>$0.44 &0.57\\
 & & & {0}$^+_4$ $\rightarrow$ {2}$^+_1$&$>$0.049 &9.8$\times 10^{-4}$\\
 & & &{7}$^-_1$ $\rightarrow$ {5}$^-_1$&0.50(5) &0.1\\
\hline
& &    &   &     &    \\
$^{204}$Pb & Expt. & SM & $^{206}$Pb  & Expt. & SM     \\
\hline
{2}$^+_1$ $\rightarrow$ {0}$^+_1$&4.69(5) &1.1 &{2}$^+_1$ $\rightarrow$ {0}$^+_1$&2.80(9) &0.9\\
4$^+_1$ $\rightarrow$ 2$^+_1$  & 0.00382(9)    & 0.1&4$^+_1$ $\rightarrow$ {2}$^+_1$&NA &0.6     \\
{0}$^+_2$ $\rightarrow$ {2}$^+_1$&0.81(25) &0.2  & {0}$^+_2$ $\rightarrow$ {2}$^+_1$&NA &3.0$\times10^{-3}$ \\
 {7}$^-_1$ $\rightarrow$ {5}$^-_1$&$\approx$0.6 &0.2 &{0}$^+_3$ $\rightarrow$ {2}$^+_1$&NA &0.2 \\
{7}$^-_1$ $\rightarrow$ {9}$^-_1$&0.15$^{+4}_{-6}$ &0.027 &    {0}$^+_4$ $\rightarrow$ {2}$^+_1$&NA &0.1\\
 
\hline

 \end{tabular}
 \end{table*}

\begin{table}        
\begin{center}
\caption{\label{qm} 
Comparison between calculated and experimental\ cite{nndc} magnetic and quadrupole moments. 
 The effective charges are taken as $e_p$ = 1.5$e$ and $e_n$ = 0.5$e$ for quadrupole moment.  The gyromagnetic ratios for magnetic moments are taken as $g_l^\nu$ = 0.00, $g_l^\pi$ = 1.00 for orbital angular momenta, and $g_s^\nu$ = -3.826, $g_s^\pi$ = 5.586 for spin angular momenta.}
\begin{tabular}{c  c c  c   c c c c}
\hline

             &   &~~~~~~~~~~~~ $\mu$($\mu_N$)  &    &~~~~~~~~~~~$Q (eb) $   &   \\
             \cline{3-4}  \cline{5-6}
     A    & $J^{\pi}$   & Expt.  &  SM   & Expt.   &  SM \\


\hline
$^{196}$Pb& $6^+_1$  &NA&  -1.20  &      NA     & -0.11 \\
         & $8^+_1$  &NA&  -1.62  &      NA     & 0.0013 \\
        & $5^-_1$  &  0.490(15)      & -0.32    &      NA       &  0.127  \\
        & $9^-_1$  &  -0.33(9)      & -0.40   &     NA      &   0.075 \\

 $^{198}$Pb & $6^+_1$  &NA&  -1.15  &      NA     & -0.135 \\
          & $8^+_1$  &NA&  -1.59  &      NA     & 0.0059 \\
         & $5^-_1$  &+0.38(3) &-0.45   &      NA     &0.21 \\
           & $7^-_1$  &   -0.377(6)      & -0.87  &    NA       &0.22   \\
         
 $^{200}$Pb     & $6^+_1$  &NA&  0.23  &      NA     & -0.11 \\
              & $8^+_1$  &NA&  -1.54  &      NA     & 0.015 \\ 
         & $7^-_1$  & -0.21(10)& -0.82  &  0.32(2)        &  0.22\\
           & $9^-_1$  &-0.257(10) &  -0.37  &      0.40(2)      &  0.24\\

 $^{202}$Pb & $4^+_1$  & +0.008(16)& 1.1  &  NA        & 0.03 \\
              & $6^+_1$  &NA&  0.48  &      NA     & 0.12 \\
              & $8^+_1$  &NA&  -1.46  &      NA     & 0.025 \\ 
             & $7^-_1$  &   NA      &-0.81   &    0.28(2)       &  0.19\\
              & $9^-_1$  &-0.2276(7) & -0.36   &      +0.58(9)      &0.32  \\
$^{204}$Pb & $2^+_1$  & $<$0.02& 0.38  &  +0.23(9)        & -0.082 \\
            & $4^+_1$  &+0.224(3) &  1.24  &      0.44(2)      & 0.16 \\
              & $6^+_1$  &NA&  1.18  &      NA     & 0.23 \\
             & $8^+_1$  &NA&  -1.55  &      NA     & 0.098 \\

$^{206}$Pb & $2^+_1$  & $<$0.030&  0.17 &  +0.05(9)        &  0.15\\
              & $6^+_1$  &NA&  -1.24  &      NA     & -0.066 \\
              & $8^+_1$  &NA&  -1.65  &      NA     & 0.017 \\
              & $6^-_1$  &+0.8(4) &  -1.86  &      NA     & 0.20 \\
              & $7^-_1$  &-0.152(3) & -0.52   &      0.33(5)      & 0.25 \\
\hline

  \end{tabular}

  \end{center}
 \end{table}

\begin{table*}
\begin{center}
\caption{The calculated half-life  of isomeric states for Pb isotopes in comparison with the experimental data (Expt.) \cite{nndc,196Pb,198Pb,200Pb,202Pb,204Pb,206Pb}.}
\label{t_hl}

\begin{tabular}{r|rcccccc}
\hline

$J^{\pi}$ & $E_{\gamma}$  & $B(E\lambda)$    &  $B(E\lambda)$  & Expt. & SM \\
& (MeV)  &    &  ($e^2$fm$^{2\lambda}$) &  T$_{1/2}$ & T$_{1/2}$ \\ 
\hline
$^{196}$Pb  & &  &  &   & &   \\
$9_1^-$ & 0.095 & $B(E2)$ & 5.90&    52(5) ns & 1.7 $\mu$s   \\

\hline
$^{198}$Pb  & &  &  &   & &   \\
$7_1^-$ & 0.129 & $B(E2)$ & 4.27&    4.19(10) $\mu$s & 1.2 $\mu$s  \\
$9_1^-$ & 0.074 & $B(E2)$ & 2.42&    137(10) ns & 4199.0 ns  \\

\hline

&   &  &   & & &      \\
$^{200}$Pb  & &  &  &   & &   \\
$7^-_1$ &  0.121 & $B(E2)$  &  0.33  &   45.2(10) ns & 17.2 $\mu$s    \\
$9^-_1$ &  0.028 & $B(E2)$  &  0.70  &   478(12) ns & 17.0 $\mu$s   \\

\hline
&   &  &   & & &    \\
$^{204}$Pb  & &  &  &   & &   \\
$4_1^+$ & 0.278 & $B(E2)$ & 8.22&    265(6) ns & 36.2 ns  \\
$7_1^-$ & 0.224 & $B(E2)$ & 1.93&    0.45$_{-3}^{+10}$ $\mu$s& 0.4 $\mu$s  \\

\hline

\end{tabular}
\end{center}
\end{table*}

The calculated results for the B(E2) values, magnetic moments, quadrupole moments and half-life of isomeric states of Pb isotopes are presented in Tables \ref{be2}, \ref{qm} and \ref{t_hl}, alongside a comparison with the corresponding experimental data.
The effective charges of protons and neutrons are assumed to be e$_p$ = 1.5e and e$_n$ = 0.5e for quadrupole moments and E2 transition rates. For magnetic moments, the gyromagnetic ratios are $g_l^\pi$ = 1.00 and $g_l^\nu$ = 0.00 for orbital angular momenta, and $g_s^\nu$ = -3.826, $g_s^\pi$ = 5.586 for spin angular momenta.
 The calculated B(E2;$0^+_1 \rightarrow 2^+_1$) value in $^{196}$Pb is small in comparison to experimental data, {\color{black} this might be due to small overlap between the wave function of $0^+$ and $2^+$ states.}
In $^{196}$Pb isotope small value for B(E2;$9^-_1 \rightarrow 7^-_1$) transition is obtained. The small $B(E2)$ value from shell model calculation 
supports isomeric character of this state.
Similarly for $^{198}$Pb, the calculated value of B(E2;$2^+_1 \rightarrow 0^+_1$) transition is small in comparison to experimental data.
The calculated, B(E2;$7^-_1 \rightarrow 9^-_1$), and B(E2;$9^-_1 \rightarrow 5^-_1$) values are small in $^{198-200}$Pb, due to which 
$7^-_1$, and $9^-$ are isomeric states in these isotopes. In $^{200}$Pb experimental data for B(E2;$2^+_1 \rightarrow 0^+_1$) is unavailable. The obtained 
B(E2;$4^+_1 \rightarrow 2^+_1$) value for $^{200}$Pb is small. 
In $^{204}$Pb, a small value for B(E2;$4^+_1 \rightarrow 2^+_1$) is obtained, due to this small value of B(E2) $4^+_1$ is an isomeric state having a half-life of {\color{black} 36.2 ns}. The $7^-_1$ state is an isomeric state in this isotope due to such a small value obtained for B(E2;$7^-_1 \rightarrow 9^-_1$) transition, and its calculated half-life comes to be {\color{black} 0.40 $\mu s$}.
In $^{206}$Pb experimental data for B(E2) is available only for B(E2;$2^+_1 \rightarrow 0^+_1$) transition and the corresponding theoretical value is underestimated by a factor of $\approx$ 3.
In Fig. \ref{be2_trend}, the comparision between calculated and experimental B(E2;2$^+_1$ $\rightarrow$ 0$^+_1$) for even-even $^{186-206}$Pb isotopes are shown, {\color{black} the comparision between calculated and experimental $2^+$ energy is also shown. The deviation in the calculated $2^+$ energy is large as we move away from shell-closure.} 
The magnetic moment and quadrupole moment for the $6^+_1$, and $8^+_1$ states for $^{196-206}$Pb are unavailable. Experimentally, the magnetic moment for $5^-_1$ state have a positive sign, whereas its theoretically calculated value have a negative sign. This $5^-_1$ state is coming due to one unpaired neutron in $p_{3/2}$, and $i_{13/2}$ each, due to which its sign for magnetic moment should be negative. The magnetic moment for $9^-_1$ state is reproduced reasonably well. Experimental data for the quadrupole moment of these states are unavailable. For $^{198}$Pb isotope the magnetic moment for $5^-_1$ state is positive experimentally, whereas its theoretically obtained value is negative. The $5^-_1$ state comes from one neutron hole in $p_{3/2}$, and $i_{13/2}$ orbital each, therefore its magnetic moment should be negative.
The magnetic moment for $7^-_1$ state is overpredicted by a factor $\approx$ 2. Experimental data for the quadrupole moment of these states are unavailable. For the $^{200}$Pb isotope theoretically calculated magnetic moment for $7^-_1$ state is overpredicted by a factor $\approx$ 4. The quadrupole moment for this state  reproduced reasonably. The magnetic moment for the $9^-_1$ state is reproduced reasonably well, whereas its quadrupole moment is underestimated by a factor $\approx$ 2. In $^{202}$Pb experimental data for {\color{black} the magnetic moment of}  $7^-_1$ state is unavailable, while its quadrupole moment shows reasonable agreement with the experimental data. The magnetic moment and quadrupole moment are reproduced reasonably well for $9^-_1$ state in $^{202}$Pb. In $^{204}$Pb the magnetic moment for the $4^+_1$ state is overpredicted almost 5 times of its experimental value, whereas quadrupole is underestimated by a factor $\approx$ 3. The theoretically calculated quadrupole moment of the $2^+_1$ state in $^{206}$Pb is three times of its experimental value. The quadrupole moment for $7^-_1$ state is reproduced reasonably well for $^{206}$Pb. The magnetic moment for $6^-_1$ state has a positive sign experimentally, whereas theoretically, its sign is negative. The $6^-_1$ state formed due to one neutron hole in $p_{1/2}$, and $i_{13/2}$ orbital each, therefore its magnetic moment should be negative.

\begin{figure}
\begin{center}
   \includegraphics[width=5.7cm,height=5cm,clip]{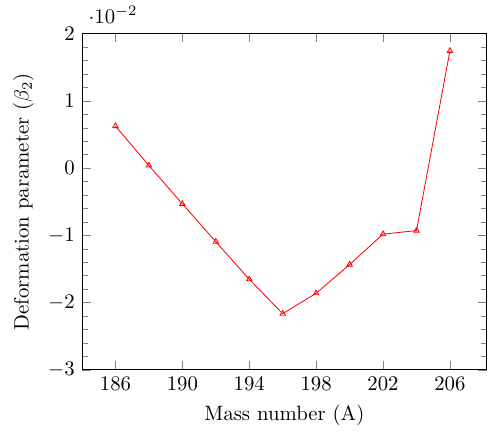}
   \caption{Shell-model results for deformation($\beta_2$).}
   \label{beta}
  \end{center} 
\end{figure}

{\color{black}
From the collective model of a nucleus, it is possible to connect between intrinsic and spectroscopic quadrupole moment, i.e. $Q_{s}$ = $<IK20|IK><II20|II>Q_{i}$, where $K$ is the projection of the nuclear spin $I$ onto the symmetry axis. With the assumption of $K = I$, we can correlate the quadrupole moment as a function of the deformation by the following expression \cite{PRL_deformation},
\begin{equation}Q_{s} = \frac {3}{\sqrt 5 \pi} <r_{0}^2> Z \beta_2 \frac {I(2I-1)}{(I+1)(2I+3)}.\end{equation}

Here, $<r_{0}^2> = (1.2 fm)^2 A^{2/3}$. In the Fig. \ref{beta}, we have shown the calculated values of deformation parameter $\beta_2$, this is a fingerprint of evolution of collectivity in the Pb isotopes with $N=104$ to $N=124$ isotopes. These values we have extracted from the calculated shell model results of Q($2_1^+$). Deformation is largest for $^{196}$Pb, there is sudden change of deformation from $^{204}$Pb to $^{206}$Pb. This is because of excitation of one neutron from $f_{5/2}$ to $p_{1/2}$ orbital. Thus one-hole configuration in $f_{5/2}$  neutron orbital is responsible for sudden  increase in the calculated quadrupole moment (i.e. 
corresponding value of $\beta_{2}$). }

\section{Summary}\label{summary}
In the present work we have performed a systematic shell model study of {\color{black} $^{196-206}$Pb} isotopes using KHH7B effective interaction. 
The results corresponding to energy spectra, electromagnetic properties and isomeric states are shown.  The results of energy spectra are in the reasonable agreement with the available experimental data. To get more accurate results it is important to perform calculation across $Z=82$ and $N=126$ shell in the extended model space.\\


\hspace{-0.6cm}{\bf Acknowledgments} We would like to thank the National Supercomputing Mission (NSM) for providing computing resources of ‘PARAM Ganga’ at the Indian Institute of Technology Roorkee.\\








\end{document}